\documentclass[11pt,a4paper,twoside]{article}
\usepackage{graphicx}
\usepackage[english]{babel}
\usepackage{color}
\usepackage{supertabular}
\usepackage[authoryear]{natbib}
\usepackage{hyperref}
\title{\normalsize{\textbf{THE RED RAIN PHENOMENON OF KERALA AND ITS POSSIBLE EXTRATERRESTRIAL ORIGIN}}}
\author{{\scshape Godfrey Louis} {and} {\scshape A. Santhosh Kumar}\\
\textit{{\small School of Pure \& Applied Physics, Mahatma Gandhi University,}}\\
\textit{{\small Kottayam-686560, India; E-mail: godfreylouis@vsnl.com}}}

\date{1 January, 2006}

\begin{document}

\oddsidemargin 0.4in
\evensidemargin 0.4in

\maketitle

\thispagestyle{myheadings}
\markright{Accepted for publication in - Astrophysics and Space Science}

\pagestyle{myheadings}
\markboth{{\small Red rain phenomenon of Kerala..}}{{\small Godfrey Louis \& Santhosh Kumar}}

\begin{abstract}
A red rain phenomenon occurred in Kerala, India starting from 25\textsuperscript{th} July 2001, in which the rainwater appeared coloured in various localized places that are spread over a few hundred kilometers in Kerala. Maximum cases were reported during the first 10 days and isolated cases were found to occur for about 2 months. The striking red colouration of the rainwater was found to be due to the suspension of microscopic red particles having the appearance of biological cells. These particles have no similarity with usual desert dust. An estimated minimum quantity of 50,000 kg of red particles has fallen from the sky through red rain.  An analysis of this strange phenomenon further shows that the conventional atmospheric transport processes like dust storms etc. cannot explain this phenomenon. The electron microscopic study of the red particles shows fine cell structure indicating their biological cell like nature. EDAX analysis shows that the major elements present in these cell like particles are carbon and oxygen.  Strangely, a test for DNA using Ethidium Bromide dye fluorescence technique indicates absence of DNA in these cells.  In the context of a suspected link between a meteor airburst event and the red rain, the possibility for the extraterrestrial origin of these particles from cometary fragments is discussed. 

\end{abstract}

\footnotetext{ The original publication will be available at www.springerlink.com after the date of publication.}

\section{Introduction}

The mysterious red rain phenomena occurred over different parts of Kerala, a State in India, starting from 25\textsuperscript{{th}}  July 2001. The news reports of this phenomenon appeared in various newspapers and other media \citep{nature01}and are currently carried by several websites \citep{rama01,radha01,surendran01,solomon01,nair01}. In an unpublished report, \citet {sampath01} claimed that the red rain particles were possibly fungal spores from trees. But they also raised several unexplained questions regarding the origin of huge quantity of red particles amounting to several tons and the unexplainable mechanism by which the red particles can reach the rain clouds etc.  From the observation of a dust layer in the atmosphere using multiwavelength LIDAR data of 24\textsuperscript{th}
 and 30\textsuperscript{{th}}  July 2001 above Thiruvananthapuram (8.33 deg N, 77 deg E),  \citet{satya04} and \citet{veera03} claimed that the dust generated from desert areas of the west Asian countries was a possible cause of the observed coloured rain. However their study did not address the cause of red rain that continued to occur in Kerala for an extended period of time. Further, the nature of the red particles, which coloured the red rain, was not examined in their study. In this paper we give a detailed account of the geographical and time distribution patterns of the red rain phenomenon of Kerala and also provide the photomicrograph study of the red particles. The possible biological nature of the red rain particles is also investigated through electron microscopy and elemental analysis. The result of the test for DNA using Ethidium Bromide dye fluorescence technique is also reported in this paper. It is also discussed how this phenomenon cannot be explained using ideas like desert dust storm activity. Considering the suspected connection of the red rain phenomenon with a meteor air burst event, it is further discussed, how the red rain phenomenon can be explained as due to the fall of fragments from a fragile cometary meteor that presumably contain a dense collection of red cells.

\begin{figure}[htb]
\centering
   \includegraphics[width=3.5in,bb=43 198 543 719,clip]{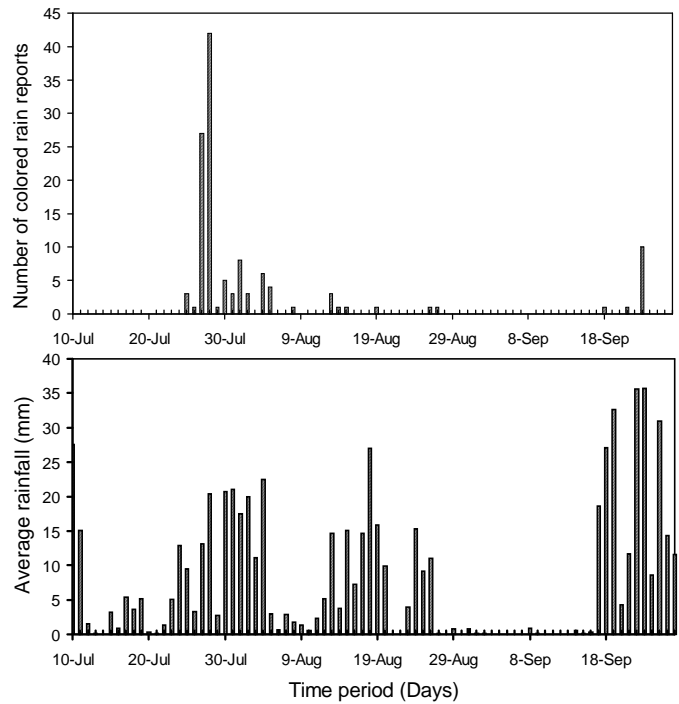}
   \caption{\small{Plot of coloured rain and rainfall data. Top panel shows a plot of the number of coloured rain incidences in Kerala on different dates.  Bottom panel shows the average of the rainfall recorded in Kerala from 10\textsuperscript{th} July to 27\textsuperscript{th} September 2001.}}
	 \label{fig:rainplot}
\end{figure}

\section{Red rain phenomenon}

The red coloured rain first occurred at Changanassery (9.47 deg N, 76.55 deg E) in Kottayam district on 25\textsuperscript{th} July 2001 and continued to occur with diminishing frequency in Kottayam and other places in Kerala for about two months.  In majority of the cases the colour of the rain was red. There were a few cases of yellow coloured rain and rare unconfirmed cases of other colours like black, green, gray etc. Coloured hailstones were also reported.

Some of the observed facts regarding this phenomenon are stated as follows. The red colour of the rainwater is due to the mixing of a particular type of microscopic red particles having the appearance of biological cells. This is confirmed by microscopic examination of the rainwater samples collected from various locations separated by more than 100 kilometers. The characteristics of the red particles contained in the red rain samples collected from different places were the same showing a common origin. The red particles were uniformly dispersed in the rainwater to impart the characteristic red colour.  The red rainwater is basically a pure suspension of red cells and is practically devoid of any dust content.  When the red rainwater was collected and kept for several hours in a vessel, the suspended particles have a tendency to settle to the bottom of the vessel causing a colour reduction for the red rainwater.  

The red rain occurred in many places during a continuing normal rain. Vessels kept in open areas clearly away from trees and house roofs also collected red rainwater. It was reported from a few places that people on the streets found their cloths stained by red raindrops. In few places the concentration of particles were so great that the rainwater appeared almost like blood.  Another characteristics of the red rain were its highly localized appearance. It usually occurs over an area of less than a square kilometer to a few square kilometers. Many times it had a sharp boundary, which means while it was raining strongly red at a place a few meters away there were no red rain. The time duration of a typical red rain was not long; usually it lasted for a few minutes to less than about 20 minutes.

\section{Distribution pattern of red rain}

A study of the distribution pattern of the red rain incidence with location and time was done for the period from July to September 2001 using the data available on this phenomenon. This data was mostly compiled from the reports that appeared in local leading Malayalam language newspapers, which have an extensive network of reporters covering all parts of Kerala. A list of coloured rain incidences with place and date is presented in Appendix I (Table \ref {tab:appendix }) . In addition to this there can be several unobserved or unreported cases. Still the available data is sufficient to show the trend and nature of this phenomenon.

\begin{figure}[htbp]
\centering
		\includegraphics[width=3.5in,bb=77 394 459 661,clip]{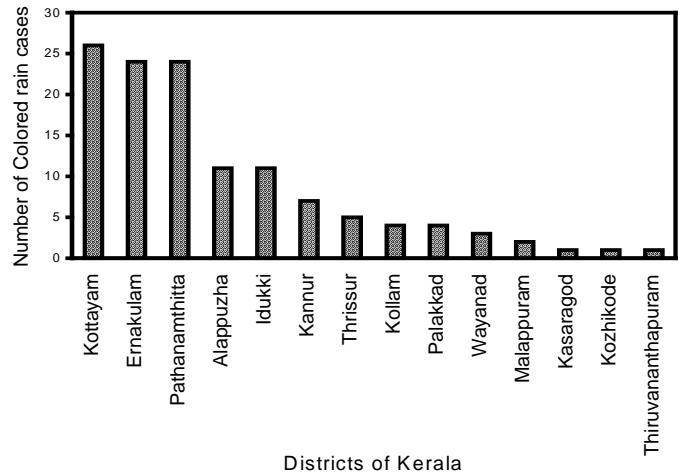}
	\caption{\small{The distribution of coloured rain phenomenon in different districts of Kerala.}}
	\label{fig:district distribution}
\end{figure}

\begin{figure}[htb]
\centering
		\includegraphics[width=3in,bb=111 143 496 639,clip]{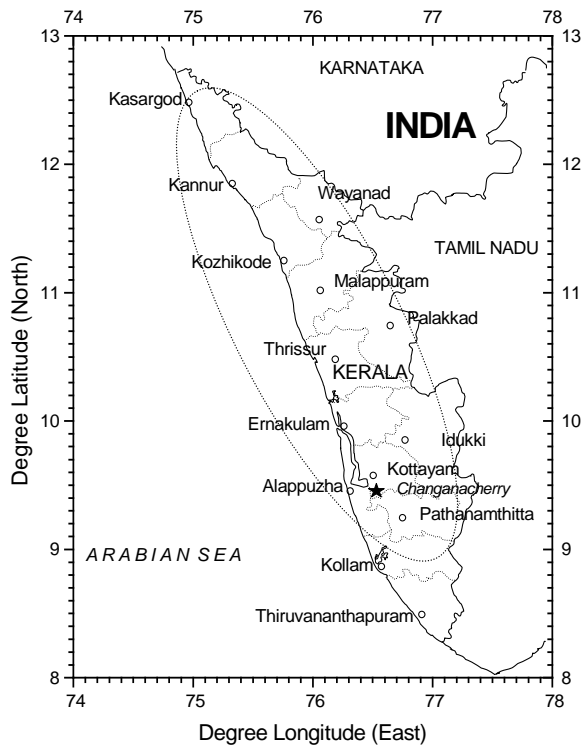}
	\caption{\small{The geographical area marked by the dotted ellipse is where the red rain incidences mainly occurred in Kerala.}}
	\label{fig:GeographicalArea}
\end{figure}

A plot (Fig.\ref{fig:rainplot} top panel) of the number of coloured rain incidences in Kerala on different dates shows that about 70\% of the total 124 listed cases occurred during the first 5 days. About 15\% of the total 124 listed cases occurred in the next 5 days and the remaining 15\% of the listed cases occurred in next 50 days in a diminishing rate. A plot (Fig.\ref{fig:rainplot} bottom panel) of the average rainfall data of Kerala enclosing the coloured rain period from 25\textsuperscript{th}  July to 23\textsuperscript{rd} September 2001, demonstrates that the coloured rain started suddenly during a period of rainfall in the state. The geographical distribution (Figs. \ref{fig:district distribution}) of the red rain cases shows a clustering of cases in Kottayam and neighboring districts like Pathanamthitta, Ernakulam, Idukki and Alappuzha with abrupt decrease towards the south and gradual decrease towards the north. The maximum numbers of cases were reported from in and around Kottayam and Pathanamthitta districts. The elliptical region marked in the map (Figure \ref{fig:GeographicalArea}) indicates the region in which red rain cases were mainly distributed.

\section{The red rain particles}

\begin{figure}[htb]
	\centering
	\includegraphics[width=3.5in]{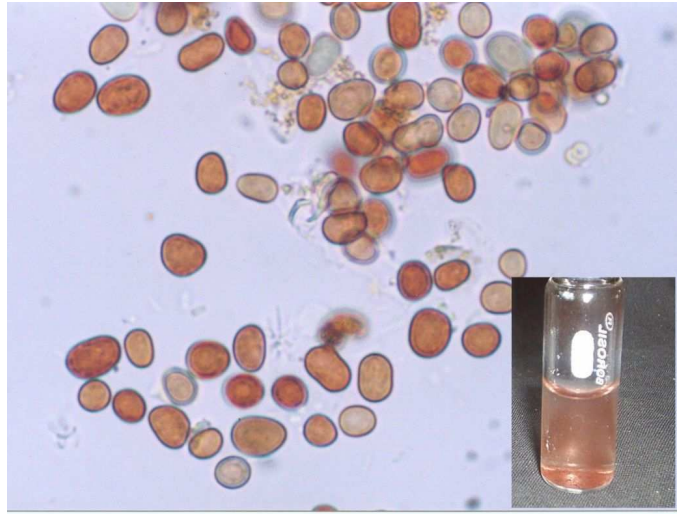}
	\caption{\small{Photomicrograph of the red rain particles under 1000x magnification.  Particles have size variation from 4 to 10 micrometers. Inset shows red rainwater contained in a 5ml sample bottle.}}
	\label{fig:Photomicrograph}
\end{figure}

The samples of red rainwater used in this study were obtained from widely different geographical locations separated by more than 100 kilometers from districts of Kottayam, Pathanamthitta and Ernakulam. A typical sample of the red rainwater is shown as inset in figure 4. The colour causing red rain particles are about 4 to 10 {\(\mu\)}m in size, almost transparent red in colour and are well dispersed in the rainwater. Under low magnification the particles look like smooth, red coloured glass beads. Under high magnifications (1000 x) their differences in size and shape can be seen (Fig. \ref{fig:Photomicrograph}). Shapes vary from spherical to ellipsoid and slightly elongated types. The particles have an appearance similar to unicellular organisms.   These cell like particles have a thick and coloured cell envelope, which can be well identified under the microscope. In a large collection only a few were found to have broken cell envelopes.  No nucleus could be observed in these cells even after staining with acidified methyl green dye. This cell like red particles clearly shows a layered structure after the dye penetration. The majority of the red rain particles have reddish brown colour under transmitted light but a small percentage of particles are white or have colours with light yellow, bluish gray and green tints. These cell-like particles do not have any flagella as found in many algae cells. The particles are very stable against decay with time. Even after storage in the original rainwater at room temperature without any preservative for about 4 years, no decay or discolouration of the particles could be found. 

\section{UV- visible absorption spectrum of red rainwater}

\begin{figure}[htb]
	\centering
	\includegraphics[width=3in,bb=57 485 485 719,clip]{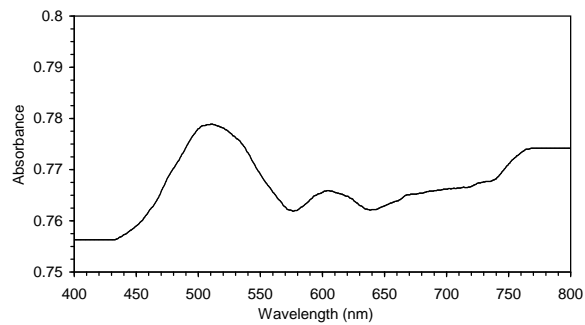}
	\caption{\small{Absorption spectra of red rainwater in the visible region}}
	\label{fig:Absorptionvisible}
\end{figure}

\begin{figure}[htb]
	\centering
		\includegraphics[width=3in,bb=27 239 488 554,clip]{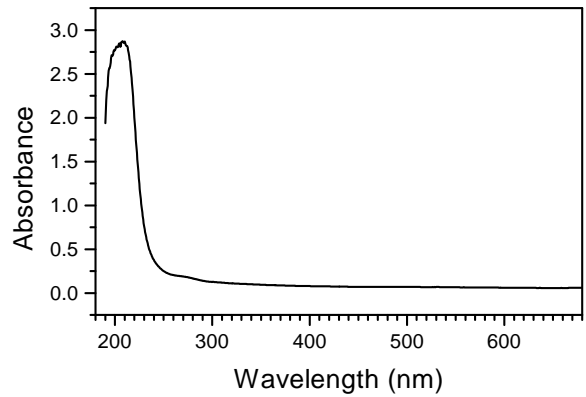}
	\caption{\small{UV absorption peak of the red rainwater near 200nm}}
	\label{fig:UVAbsorption}
\end{figure}

The absorption spectrum of the red rainwater in the visible region was recorded using Spectrophotometer (Shimadzu Model No. UV-2401 PC). The spectrum (Fig. \ref{fig:Absorptionvisible}) shows a major absorption peak at 505 nm and a minor peak at 600 nm. The red colouration is clearly due to the absorption of the green and yellow wavelengths by the 505 nm absorption peak. There is a reduction in the absorption towards the blue end. This is the reason for the slight pink tint of the red rainwater. The UV- Visible absorption spectrum of the dilute red rainwater is shown in Figure \ref{fig:UVAbsorption}. There is a clear absorption peak near 200 nm.  Features of the visible region are negligibly small when compared to the large absorption peak near 200nm. 

\section{Number density and total mass of red cells}

The number density of the red rain particles in the rainwater was found using a counting chamber under the microscope. The average number density was found to be 9x10\textsuperscript{6\ }particles per 1ml of rainwater. This figure had a variation of about 30\% with samples from various locations. 

A quantity of 30 ml of a typical sample of red rainwater was dried and the weight of the dried red particles was determined using a microbalance. From this, the approximate weight of the particles contained in one litre of rainwater was found to be 100 mg or in 1 cubic meter the weight of red particles is 0.1 kg. Consider a typical case of red rain where a minimum 5 mm of red rain has fallen in an area of 1 sq. km. This works out to 5000 cubic meters of water and the weight of the red particles contained in it will be 500 kg. There are more than 100 reported cases of red rain, which means more than 50000 kg of red particles are involved in this phenomenon. This is a minimum estimate and the weight of the total red particles can be much more if several unnoticed and unreported cases are assumed. Further the red rain was more intense and widespread in many locations than assumed in the typical case. It is a mystery from where the rain clouds have picked up such a large quantity of pure red particles.   

\begin{figure}[htb]
	\centering
	\includegraphics[width=2.5in]{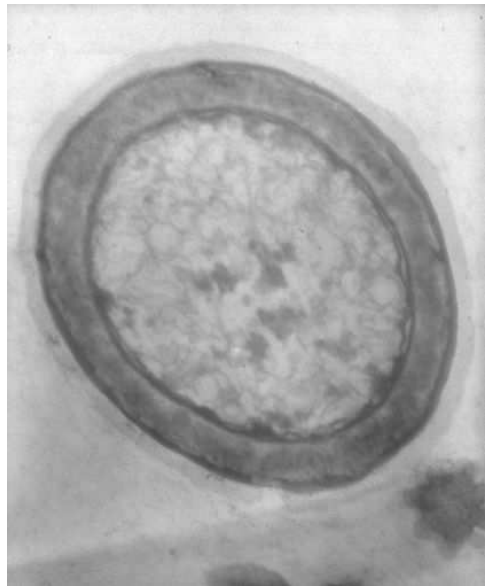}
	\caption{\small{TEM image of a typical red rain cell under 25000x magnification showing the thick outer cell envelope}}
	\label{fig:TEMImage7}
\end{figure}

\begin{figure}[htb]
	\centering
	\includegraphics[width=2.5in]{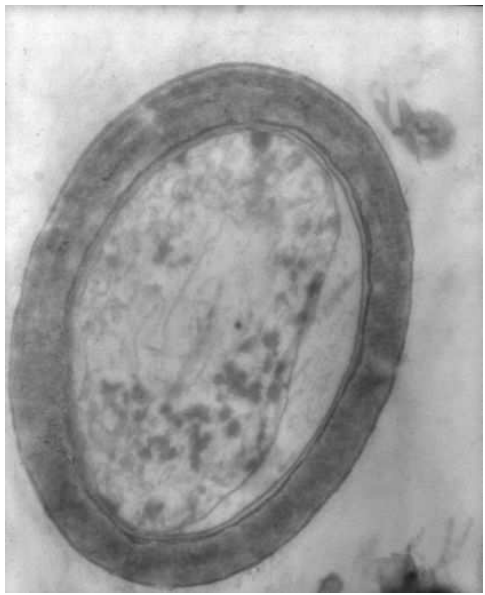}
	\caption{\small{TEM image of an oval shaped red rain cell under 20000x magnification showing a detached inner capsule.}}
	\label{fig:TEMImage8}
\end{figure}

\begin{figure}[htb]
	\centering
	\includegraphics[width=2.5in]{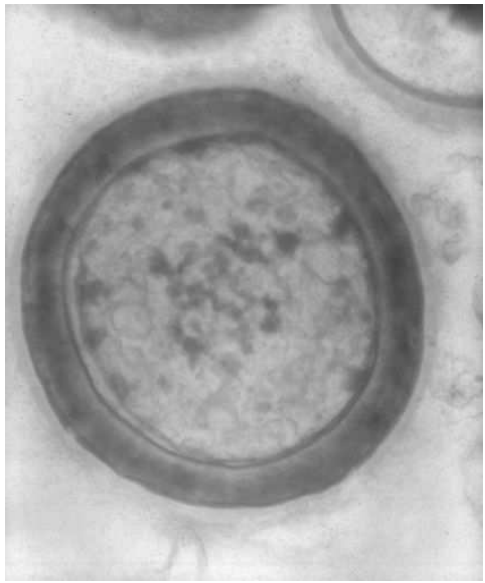}
	\caption{\small{TEM image of another typical red rain cell under 20000x magnification. }}
	\label{fig:TEMImage9}
\end{figure}

\begin{figure}[htb]
	\centering
	\includegraphics[width=2.5in]{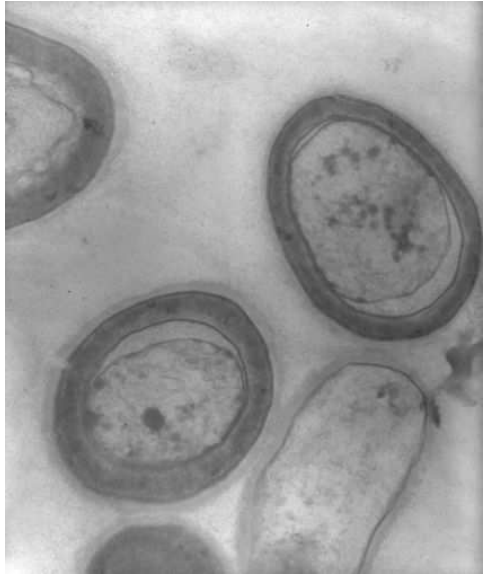}
	\caption{\small{TEM image of a collection of red rain cells under 10000x magnification. }}
	\label{fig:TEMImage10}
\end{figure}

\begin{figure}[htb]
	\centering
	\includegraphics[width=2.5in]{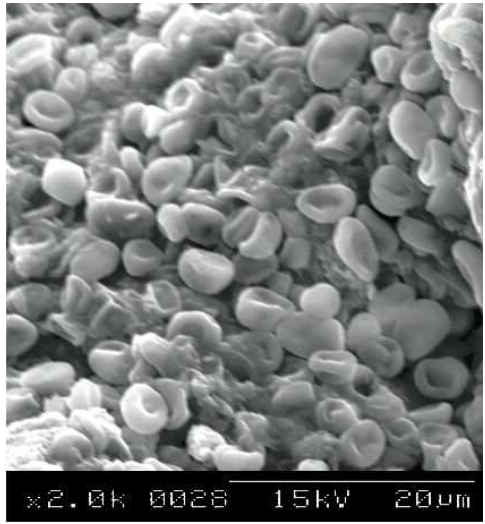}
	\caption{\small{ SEM image of a cluster of red rain cells (magnification 2000x).}}
	\label{fig:SEMImage11}
\end{figure}

\begin{figure}[htb]
	\centering
	\includegraphics[width=2.5in]{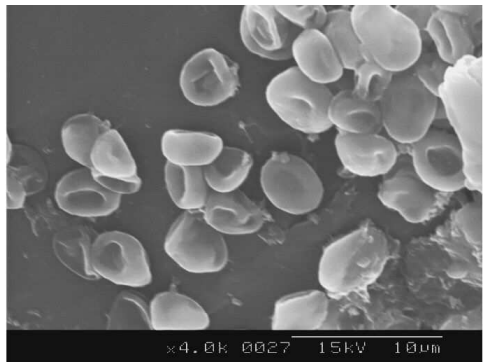}
	\caption{ \small{SEM image of red rain cells under 4000x magnification showing the squeezed appearance.}}
	\label{fig:SEMImage12}
\end{figure}

\begin{figure}[htb]
	\centering
	\includegraphics[width=2.5in]{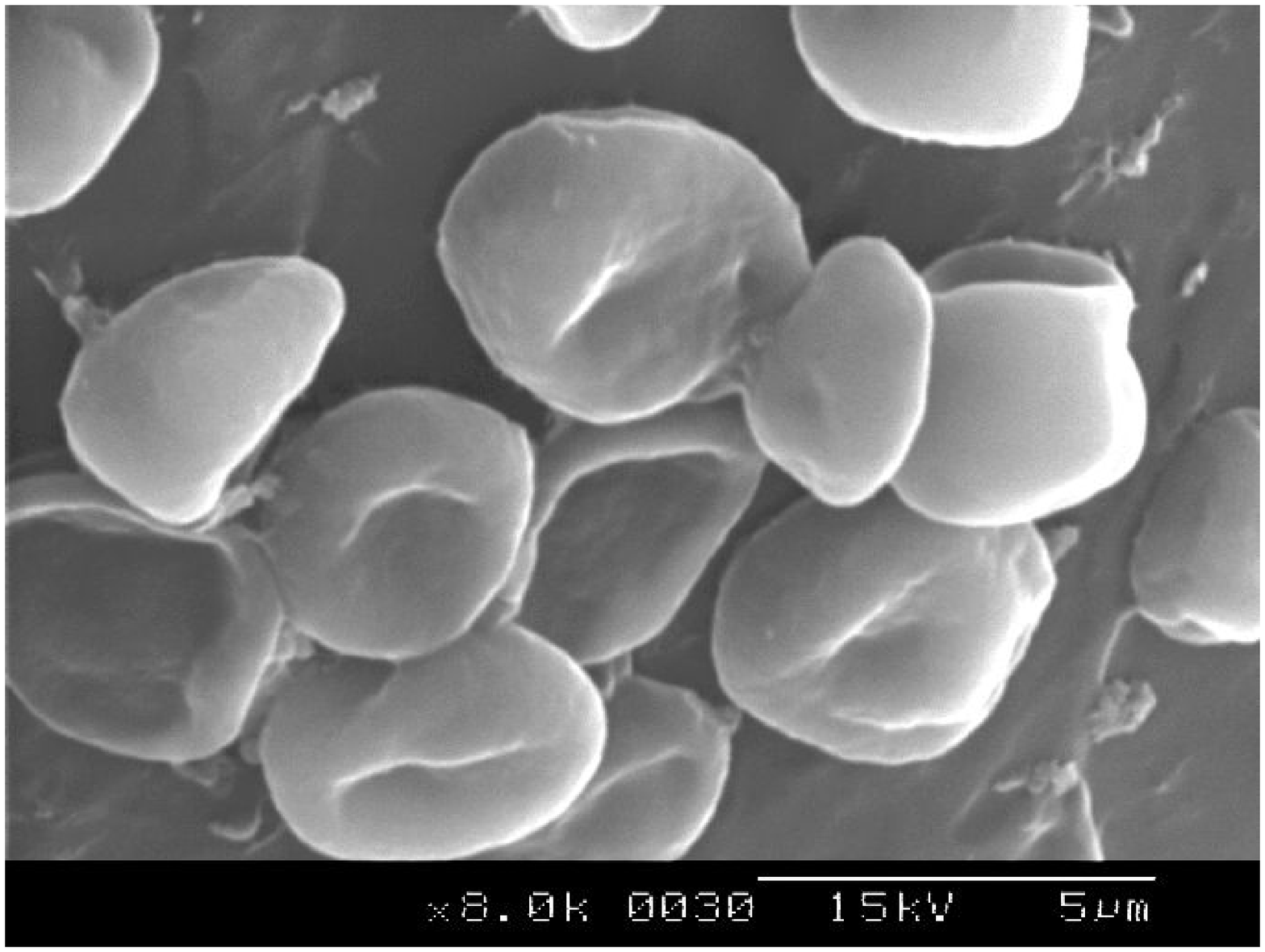}
	\caption{\small{SEM image of red rain cells under high magnification (8000x) showing some fine surface features. }}
	\label{fig:SEMImage13}
\end{figure}

\section{ Electron microscopy study}

The TEM images were taken using Hitachi model H600 electron microscope.   Examination of the TEM images shown in figures \ref {fig:TEMImage7}, \ref{fig:TEMImage8}, \ref {fig:TEMImage9} and \ref {fig:TEMImage10} clearly shows that these particles are having a fine structure similar to biological cells. The images show that these cells do not have a nucleus.  Cell wall is comparatively thick. The images show that these red rain particles have fine-structured membranes.  Encased inside the thick outer wall there appears to be a detached inner capsule, which contains the cells inner substance. In some region the inner capsule appears to be detached from the outer wall to form an empty region inside the cell (fig.\ref{fig:TEMImage8}). Further there appears to be a faintly visible mucus layer present on the outer side of the cell. 

The Scanning Electron microscope images are shown in Figures \ref{fig:SEMImage11}, \ref{fig:SEMImage12} and \ref {fig:SEMImage13}. The outer surface of these particles is smooth and round (fig.\ref {fig:SEMImage13}). One characteristic feature is the inward depression of the spherical surface to form cup like structures giving a squeezed appearance (fig.\ref{fig:SEMImage12}). The amount of such surface deformation varies from cell to cell and some of the cells do not have these surface depressions.  As found in the optical microscopy and TEM, there are no flagella or filamentous structures attached to the outer surface these cells. The cell size varies from 4 to 8 micrometers.

\begin{figure}[htb]
	\centering
	\includegraphics[width=3in, bb=115 301 488 612, clip]{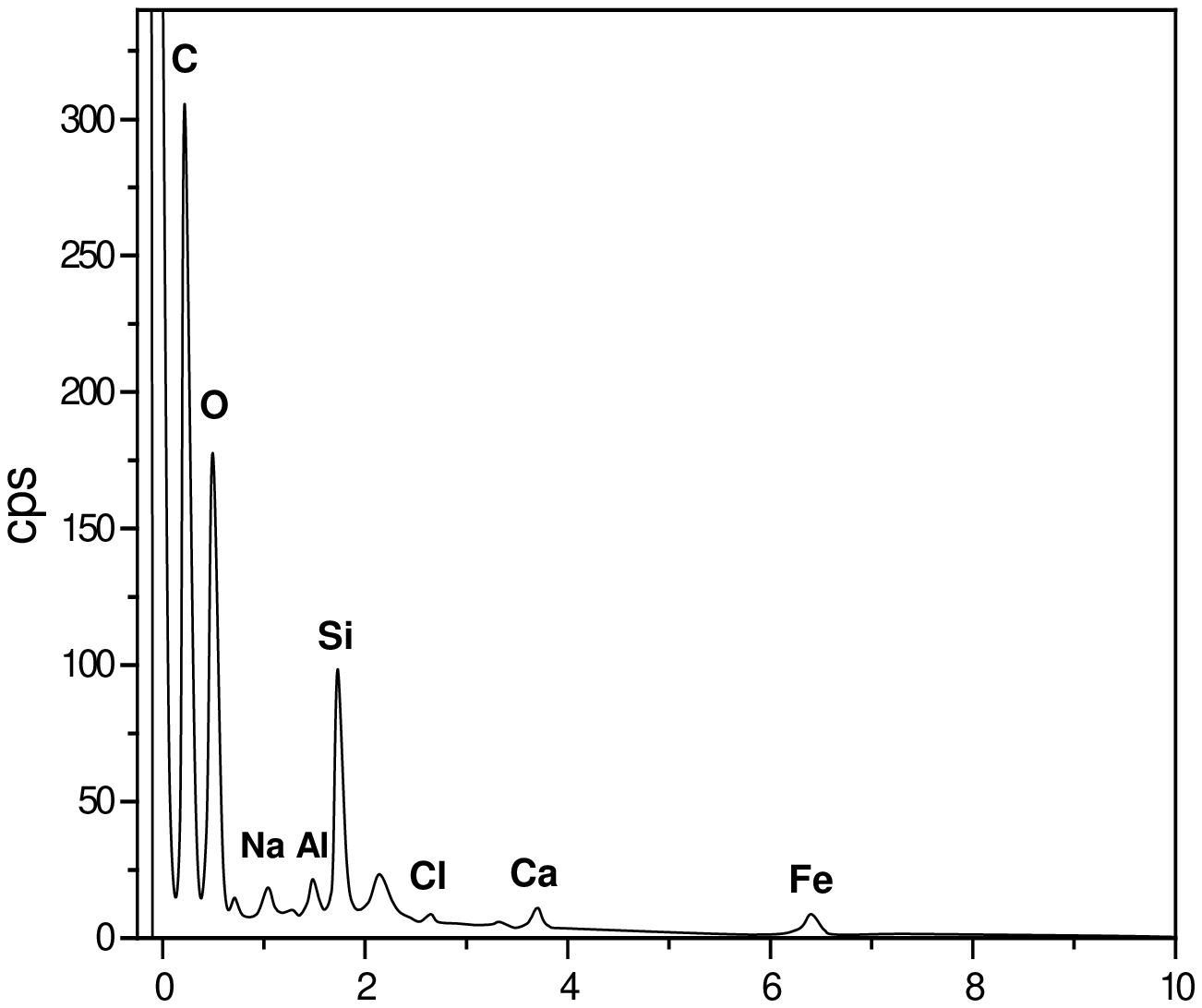}
	\caption{\small{Energy Dispersive X-ray Spectrum (EDAX) of red rain cells showing the elemental composition. }}
	\label{fig:EDAX}
\end{figure}

\section{Elemental analysis using EDAX and CHN analyzer}

The elemental composition of the red particles was determined using the SEM attached with the energy-dispersive X-ray analyzer system (EDAX) (Hitachi Scanning electron microscope). The microscope was operated at an acceleration voltage of 9.7 KeV and in the magnification between 5000 and 8000. X-ray spectrum was recorded from an area that circumscribe the specimen. The EDAX spectrum of the red particle is shown in figure \ref{fig:EDAX} and table  \ref{tab:ElementalComposition} shows the percentage composition of the detected elements. The major constituents of the red particles are carbon and oxygen. Silicon is most prominent among the minor constituents, which includes Fe, Na, Al and Cl. 

\begin{table}
	\centering
		
			\begin{tabular}{llll}
\hline

{Element}& {Wt \%}&{Atomic \%}&{Standards}\\
\hline
{C}&{49.53}&{57.83}&{CaCO$_{3}$}\\

{O}&{45.42}&{39.82}&{Quartz}\\

{Na}&{0.69}&{0.42}&{Albite}\\

{Al}&{0.41}&{0.21}&{Al$_{2}$O$_{3}$}\\

{Si}&{2.85}&{1.42}&{Quartz}\\

{Cl}&{0.12}&{0.05}&{KCl}\\

{Fe}&{0.97}&{0.24}&{Fe}\\
\hline
\end{tabular}
		
	\caption{\small{Elemental composition of red cells by EDAX analysis}}
	\label{tab:ElementalComposition}
\end{table}

The elemental composition of the red cells was further checked using a CHN analyzer (Model Elementar Vario EL III). The presence of carbon, hydrogen and nitrogen can be analyzed using this analyzer. About 30 ml of red rainwater when dried gave a solid residue of about 3mg. This under CHN analysis showed 43.03\% carbon, 4.43\% hydrogen and 1.84\%  nitrogen.

\section{Test for DNA and RNA }

The test for the DNA and RNA is performed by spectrofluorimetric technique using ethidium bromide fluorescent dye. This dye has the property of greatly enhancing its fluorescence emission in the presence of DNA or RNA. For performing the test the cells were centrifuged out and were well crushed in a mortar for exposing the inner contents of the cells to the dye solution. 100 {\(\mu\)L of this crushed cell suspension was added to 5 ml of ethidium bromide stock solution. Fluorescence emission of this mixture at 600 nm was recorded using spectrofluorimeter (RF-5301 PC, Shimadzu) using an excitation wavelength 530 nm. This spectrum (Fig. \ref {fig:DNA}b) does not show an enhanced fluorescence in comparison with the pure ethidium bromide solution (Fig. \ref{fig:DNA}a), thus indicating the absence of DNA or RNA. This experiment was repeated after grinding the cells in liquid nitrogen to further ensure the cracking of cells.  This also shows no enhancement in fluorescence. Similar experiments when performed on a suspension of yeast cells (quantity 100 }{\(\mu\)}{L) showed greatly enhanced fluorescence effect indicating the presence of DNA (Fig.\ref{fig:DNA}c). Similarly other DNA containing plant materials also caused enhanced fluorescence confirming the validity of the test. 

\begin{figure}[htb]
	\centering
	\includegraphics[width=3in,bb=66 498 524 745, clip]{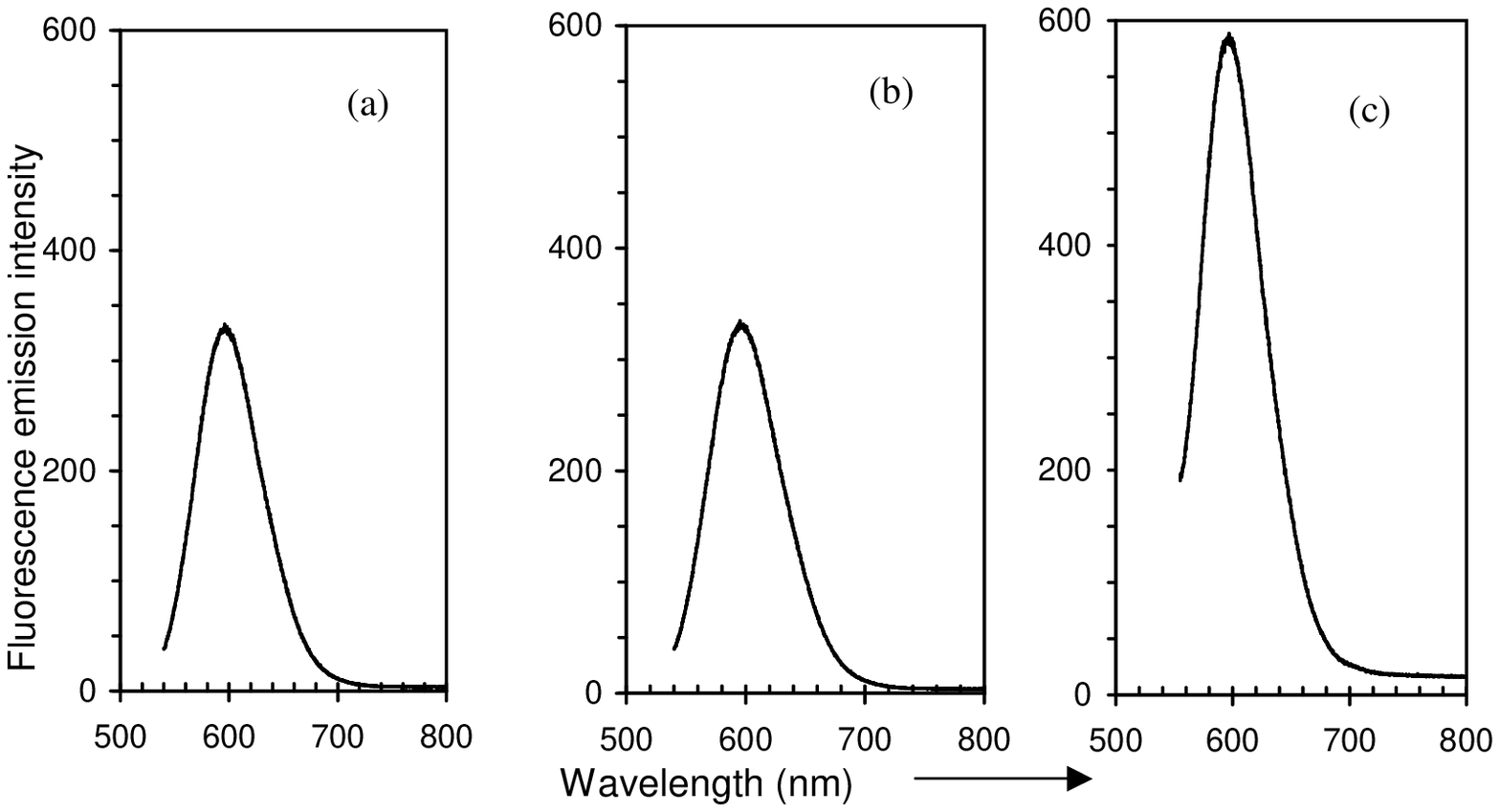}
	\caption {\small{Fluorescence emission spectrum of ethidium bromide (EtBr) dye soluition a) pure EtBr solution b) with red rain cells extract added to EtBr solution c) with yeast cell extract added to EtBr solution. }}
	\label{fig:DNA}
\end{figure}

\section{ Discussion}

When the red rain reports are viewed in the background of the normal rainfall data the pattern that emerges is that of a sudden starting of red rain phenomenon after 25\textsuperscript{th} July 2001 and then a decay of red rain cases with time. The red rain started in the State during a period of normal rain, which indicate that the red particles are not something, which accumulated in the atmosphere during a dry period and washed down on a first rain. It was found that several cases of red rain phenomenon have occurred on rainy days after and during normal rains. Thus it cannot be again assumed that the red particles came from accumulation in the lower atmosphere. The vessels kept in open space also collected red rain. Thus it is not something that is washed out from rooftops or tree leaves. Considering the huge quantity of red particles fallen over a wide geographic area, it is impossible to imagine that these are some pollen or fungal spores which have originated from trees.

The nature of the red particles rules out the possibility that these are dust particles from a distant desert source. These red particles do not have any similarity with the usual desert dust. This is clearly shown by microscopic study of the particles. Particles of this type are not found in Kerala or nearby place. The origin of these particles is unknown. It is convenient to assume that these particles are something, which got airlifted from a distant source on Earth by some wind system. Several questions remain unanswered even under such an assumption. One characteristics of each red rain case is its highly localized appearance. If particles originate from distant desert source then why there were no mixing and thinning out of the particle collection during transport. Why some isolated cases of red rain occurred over an extended period of two months despite the changes in climatic conditions and wind pattern spanning over two months. It is also unexplainable why there is a concentration of red rain incidences in Kottayam and nearby districts. Above arguments and facts indicate that it is difficult to explain the red rain phenomenon by using usual arguments like dust storms etc.

An examination of the several characteristics of this red rain phenomenon shows that it is possible to explain this by assuming the meteoric origin of the red particles. The red rain phenomenon first started in Kerala after a meteor airburst event, which occurred on 25\textsuperscript{th} July 2001 near Changanacherry in Kottayam district. This meteor airburst is evidenced by the sonic boom experienced by several people during early morning of that day. The first case of red rain occurred in this area few hours after the airburst event. This points to a possible link between the meteor and red rain. If particle clouds are created in the atmosphere by the fragmentation and disintegration of a special kind of fragile cometary meteor that presumably contain a dense collection of red particles, then clouds of such particles can mix with the rain clouds to cause red rain. The atmospheric fragmentation of the fragile cometary meteor can be the reason for the geographical distribution of the red rain cases in an elliptical area of size 450 km by 150 km. Maximum cases of red rain occurred in Kottayam and nearby districts (fig. \ref{fig:GeographicalArea}). From this, it can be inferred that while falling to the ground at low angle, the meteor has been travelling from north to south in a south-east direction above Kerala with a final airburst above Kottayam district. During its travel in the atmosphere it must have released several small fragments, which caused the deposition of cell clusters in the atmosphere from north to south above Kerala.

An examination of the red rain data shows that more than 85\% of the red rain cases occurred during the first 10 days after the airburst event. This delayed time distribution for the first few days can be accounted as due to the slow settling of the microscopic red rain particles in the atmosphere, with a settling rate of a few hundred meters per day.  For this the meteor disintegration is expected to provide a vertical distribution of particles spanning over a few kilometres above the rain clouds.  The remaining 15 \% of the isolated delayed red rain cases occurred with a delay of up to 60 days, which presumably also reflect gradual settling of the particles in the upper atmosphere. 

The biological cell like nature of the red rain particles is revealed by the electron microscopy and elemental analysis. Fine structure and enclosing cellular membranes in the red rain particles as evidenced by TEM is indicative of biological-like cells.  The external morphology of the cells as reveled by the SEM is also indicating that the red particles are like biological cells. The optical microscope images also support the idea that these transparent red particles are similar to biological cells. The clear presence of carbon as shown by the elemental analysis indicates the organic nature of these particles. While these particles have striking morphological similarity with biological cells, the test for DNA gives a negative result, which argues against their biological nature.

	The present study of red rain phenomenon of Kerala shows that the particles, which caused the red colouration of the red rain, are not possibly of terrestrial origin. It appears that these particles may have originated from the atmospheric disintegration of cometatory meteor fragments, which are presumably containing dense collections of red rain particles.  These particles have much similarity with biological cells though they are devoid of DNA. Are these cell like particles a kind of alternate life from space? If the red rain particles are biological cells and are of cometary origin, then this phenomena can be a case of cometary panspermia \citep{hoyle99} were comets can breed microorganisms in their radiogenically heated interiors and can act as vehicles for spreading life in the universe. Future collaborative studies are expected to provide more answers.

\section* {Acknowledgements}

We greatly acknowledge the help of Dr. George Varughese for collecting many of the red rain samples and important information regarding the phenomenon. We also thank Dr. A. M. Thomas for first approaching us with a coloured rainwater sample.  Thanks are also due to several others who have helped for the sample collection and provided related information about the phenomenon. We greatly acknowledge the services of Central Marine Fisheries Research Institute, Cochin, for the TEM studies, Sree Chitra Tirunal Institute for Medical Sciences and Technology, Thiruvananthapuram for the SEM-EDS studies and Sophisticated Test and Instrumentation Centre, CUSAT, Cochin for the CHN analysis.

\section*{Appendix I}

\begin {footnotesize}
\centering
\tablecaption{\small {A List of 124 reports of colored rain cases in Kerala during the period July \textendash September 2001, compiled from various news reports and other sources (List sorted by district)}}
\label{tab:appendix } 
\tablefirsthead{%
\hline
No.& District& Place& Location& Date (M/D/Y)& Description\\
\hline}

\tablehead{%
\multicolumn{6}{l}{\small\sl continued from previous page}\\
\hline
No.& District& Place& Location& Date (M/D/Y)& Description\\
\hline}

\tabletail{%
\hline
\multicolumn{6}{r}{\small\sl continued on next page}\\}

\tablelasttail{\hline

\hline
\multicolumn{6}{r}{\small\sl end of table}\\}

\begin{supertabular}{p{0.5cm}p{2.25cm}p{2.25cm}p{2.25cm}p{1.25cm}p{1.25cm}}
\hline

1& Alappuzha& Koyana& & 7/27/01&  red\\
  
2& Alappuzha& Chengannur& & 7/27/01&  red\\
 
3& Alappuzha& Pandanadu& & 7/27/01&  red\\
 
4&Alappuzha&Mulakkuzha&&7/27/01& red\\

5&Alappuzha&Pavukkara&&7/27/01& red\\
 
6&Alappuzha&Kuttamperur&&7/27/01& red\\
 
7&Alappuzha&Aroor&&7/27/01& red\\

8&Alappuzha&Harippad&&7/27/01& red\\
 
9&Alappuzha&Harippad&Thulamparambu&7/28/01& red\\
 
10&Alappuzha&Mavelikkara&Cheukol&7/28/01& red\\
 
11&Alappuzha&Mavelikkara&Ponnezha&7/28/01& red\\
 
12&Ernakulam&Perumbavur&Nedunghappra&7/27/01&black~and  red\\
 
13&Ernakulam&Edayar&Kaintekara to Kadungalloor&7/27/01& red\\
 
14&Ernakulam&Mukkannur&&7/27/01& red\\

15&Ernakulam&Kottuvalli&&7/27/01& red\\

16&Ernakulam&Parur&Kadungallur&7/27/01& red\\

17&Ernakulam&Muvattupuzha&Puthuppadi&7/27/01& red\\

18&Ernakulam&Nedumbassery&Avanamkodu&7/27/01& red\\

19&Ernakulam&Kaitharam&&7/27/01& red\\

20&Ernakulam&Alwaye&Chenganodu&7/27/01& red\\
 
21&Ernakulam&Muvattupuzha&Mazhuvannur&7/28/01&yellow\\

22&Ernakulam&Alwaye&Edayar, Panayikulam, Valluvalli, Kongorppally&7/28/01& red\\
 
23&Ernakulam&Perumbavur&Vazhakkulam&7/28/01& red\\

24&Ernakulam&Nedumbassery&&7/28/01& red ~and yellow\\
 
25&Ernakulam&Chengamanadu&&7/28/01& red\\
 
26&Ernakulam&Kochi&Fortkochi&7/28/01& red\\
 
27&Ernakulam&Kochi&Mattamcherri &7/28/01& red\\
 
28&Ernakulam&Kochi&Pallurithi&7/28/01& red\\
 
29&Ernakulam&Mulamthurithi&Karavatte&7/28/01& red\\
 
30&Ernakulam&Kezhmadu&Chundi, Chunangamveli&7/28/01& red\\
 
31&Ernakulam&Kochi&Kadavanthra&8/8/01& red\\
 
32&Ernakulam&Nedumbassery&Chengamnad, Kulavankunnu&8/13/01& red\\
 
33&Ernakulam&Angamali&Aiyampuzha, Karukutty&8/13/01& red\\
 
34&Ernakulam&Malayattoor&Thottukavala&8/13/01& red\\
 
35& Ernakulam& Kochi& Ponnurunthi& 9/18/01&  red\\
 
36&Idukki&Vannappuram&Killippara&7/25/01& red\\
 
37&Idukki&Adimali&Machiplavu&7/28/01&yellow ~and  red\\
 
38&Idukki&Kanjar&&7/28/01& red\\
 
39&Idukki&Thodupuzha&&7/28/01&yellow\\
 
40&Idukki&Nariyampara&&7/28/01& red\\
 
41&Idukki&Deviyar&&7/28/01& red\\
 
42&Idukki&Vellathuval&Panar&7/31/01& red\\
 
43&Idukki&Erattayar&Natthukallu&8/1/01& red\\
 
44&Idukki&Kattappana&Vallakkadavu&8/1/01& red\\
 
45&Idukki&Kattappana&&8/2/01&black ~and  red\\
 
46&Idukki&Kattappana&Irattayar&8/4/01& red\\
 
47&Kannur&Kuttuparambu&Thokkilangadi&7/26/01&light  red\\
 
48&Kannur&Peringom&&7/31/01& red\\
 
49&Kannur&Panoor&Chendayat&9/21/01& red\\
 
50&Kannur&Pallikunnu&Ramatheru, Kanathur &9/23/01& red\\
 
51&Kannur&Pallikunnu&Puthiyatheru&9/23/01& red\\
 
52&Kannur&Edakkad&Koshormoola&9/23/01&red\\

53&Kannur&Pappinisseri&Keecheri&9/23/01& red\\
 
54&Kasaragode&Madikai&Malappacherry&8/1/01& red\\

55&Kollam&Sooranad&&7/28/01& red\\
   
56&Kollam&Pathanapuram&&7/28/01&red\\
 
57& Kollam& Mavila& &8/26/01&\\
 
58&Kollam&Pattazhi&&8/27/01&\\
 
59&Kottayam&Changanacherry&Morkulangara&7/25/01& red\\
 
60&Kottayam&Changanacherry&Puzhavathu&7/25/01&\\
 
61&Kottayam&Puthuppally&Thrikkothaman\-galam&7/27/01& red\\
 
62&Kottayam&Kanjirappalli&Chenappady&7/27/01& red\\
 
63&Kottayam&Pala&Kadanadu&7/27/01& red, white turbid\\
 
64&Kottayam&Karukachal&Edayappara&7/28/01&deep yellow \\
 
65&Kottayam&Changanacherry&Poovam&7/28/01& red\\
 
66&Kottayam&Pambadi&Vazhoor&7/28/01& red\\
 
67&Kottayam&Ammancherri&&7/28/01&yellow\\
 
68&Kottayam&Neendoor&Pravattom&7/28/01& red\\
 
69&Kottayam&Changanacherry&Vazhappally&7/28/01& red\\
 
70&Kottayam&Vechoochira&Mannadishala&7/30/01& red\\
 
71&Kottayam&Ayirur&Edappavoor&7/30/01& red\\
 
72&Kottayam&&Planghaman&7/30/01& red\\
 
73&Kottayam&&Pullappram&7/30/01& red\\
 
74&Kottayam&Manimala&Market Jn.&8/2/01&yellow\\
 
75&Kottayam&Elikkulam Panchayat&Panamattom&8/2/01& red\\
 
76&Kottayam&Manimala&Mini Indust. Estate&8/4/01& red\\
 
77&Kottayam&Manimala&&8/5/01& red\\
 
78&Kottayam&Erumali&Thumarampara&8/5/01&yellow\\
 
79&Kottayam&Mundakkayam&Punchavayal&8/5/01&yellow\\
 
80&Kottayam&Changanacherry&Morkulangara&8/14/01& red\\
 
81&Kottayam&Elikkulam Panchayat&Panamattom&8/15/01&\\
 
82&Kottayam&Ettumanoor&Pattithanam&8/19/01& red ~and yellow\\
 
83&Kottayam&Kanjirappalli&Kappad&9/23/01& red\\
 
84&Kottayam&Erumeli&Erumeli&9/23/01&\\
 
85&Kozhikode&Malapparambu&&8/4/01& red\\
 
86&Malappuram&Vengara&&7/28/01&\\
 
87&Malappuram&Thirur&Menathur angadi&8/4/01&blue\\
 
88&Palakkad&Ongallur&Vadanamkurichy&7/27/01& red\\
 
89&Palakkad&Kizhakkancheri&&7/27/01& red\\
 
90&Palakkad&Vadakkancheri&&7/27/01& red\\
 
91&Palakkad&Kannambra&&7/27/01& red\\
 
92&Pathanamthitta&Valamchuzhi&&7/27/01&light  red\\
 
93&Pathanamthitta&Valamchuzhi&&7/27/01& red\\
 
94&Pathanamthitta&Chittar&Aarattukudukka&7/27/01&yellow\\
 
95&Pathanamthitta&Karimpanakuzhi&&7/28/01& red\\
 
96&Pathanamthitta&Kumbuzha&&7/28/01& red\\
 
97&Pathanamthitta&Chennerkkara&Murippara&7/28/01& red\\
 
98&Pathanamthitta&Adoor&Kilivayal&7/28/01& red\\
 
99&Pathanamthitta&Adoor&Munnalam&7/28/01& red\\
 
100&Pathanamthitta&Adoor&Karuvatta&7/28/01& red\\
 
101&Pathanamthitta&Kidangur&Kotta&7/28/01& red\\
 
102&Pathanamthitta&Panthalam&Kulanda&7/28/01& red\\
 
103&Pathanamthitta&&Kaippuzha&7/28/01& red\\
 
104&Pathanamthitta&Panthalam&Kadakkadu&7/28/01& red\\
 
105&Pathanamthitta&&Mangaram&7/28/01& red\\
 
106&Pathanamthitta&&Vallikkod&7/28/01& red\\
 
107&Pathanamthitta&&Thrikkovil&7/28/01& red\\
 
108&Pathanamthitta&Thiruvalla&Kuttur&7/28/01& red ~and yellow hailstones\\
  
109&Pathanamthitta&Thiruvalla&Peringara&7/28/01& red\\
 
110&Pathanamthitta&Ranni&Palachuvadu&7/29/01& red ~and yellow\\
 
111&Pathanamthitta&Murinjakal&Mlamthadam&8/1/01& red\\
 
112&Pathanamthitta&Kozhencherrry&&8/1/01&black  \\
 
113&Pathanamthitta&Vallikode&Vellappara Colony&8/1/01&\\
 
114&Pathanamthitta&Kadammanitta&&8/1/01&\\
 
115&Pathanamthitta&Ranni&Vechoochira&8/4/01&\\
 
116&Thiruvanantha\-puram&Vithura&Kallar&7/31/01& red\\
 
117&Trichur&Ponkunnam&Elangulam&7/30/01&light black and  red \\
 
118&Trichur&Gruvayoor&Mammiyoor&9/23/01& red\\
 
119&Trichur&Anthikad&&9/23/01&\\
 
120&Trichur&Irinjalakuda&Arippalam&9/23/01& red\\
 
121&Trichur&Puttanpedika&Tattaadi&9/23/01&pink\\
 
122&Wayanad&Padinharathara&Dam site&8/1/01& red\\
 
123&Wayanad&Ambalavayl&Edakkal&8/4/01& red\\
 
124&Wayanad&Thalappuzha&Kannothmala&8/5/01& red\\
 
\end{supertabular}

\end {footnotesize}

\end{document}